\documentclass[11pt,showpacs,preprintnumbers,superscriptaddress,amsmath,amssymb,nofootinbib]{revtex4}
\usepackage{graphicx}
\usepackage{dcolumn}
\usepackage{bm}
\usepackage{amssymb}
\usepackage{amsmath}
\usepackage{epsfig}    
\usepackage{color}
\usepackage{slashed}
\usepackage{float}
\newcolumntype{I}{!{\vrule width 1.3pt}}

\def\be{\begin{equation}}
\def\ee{\end{equation}}
\newcommand{\bea}{\begin{eqnarray}}
\newcommand{\eea}{\end{eqnarray}}
\newcommand{\nn}{\nonumber}

\begin{document}
\title{Three-loop Neutrino Mass Model with Doubly Charged Particles from Iso-Doublets}
\author{Hiroshi Okada}
\affiliation{School of Physics, KIAS, Seoul 130-722, Korea}
\affiliation{Physics Division, National Center for Theoretical Sciences, Hsinchu, Taiwan 300}
\author{Kei Yagyu}
\affiliation{School of Physics and Astronomy, University of Southampton, Southampton, SO17 1BJ, United Kingdom}

\begin{abstract}

We propose a new type of a three-loop induced neutrino mass model with dark matter candidates which are required for the neutrino mass generation. 
The smallness of neutrino masses can be naturally explained without introducing super heavy particles, namely, much heavier than a TeV scale
and quite small couplings as compared to the gauge couplings.  
We find that as a bonus, the anomaly of the muon anomalous magnetic moment can simultaneously be 
explained by loop effects of new particles. 
In our model, there are doubly charged scalar bosons and leptons from isospin doublet fields which give characteristic collider signatures. 
In particular, the doubly charged scalar bosons can decay into the same-sign dilepton  
with its chirality of both right-handed or left- and right-handed. 
This can be a smoking gun signature to identify our model and be useful to distinguish other models with doubly charged scalar bosons at collider experiments. 

\end{abstract}

\maketitle

\section{Introduction}

One of the most important questions in the standard model (SM) for elementary particle physics is why 
neutrinos have non-zero but quite small masses as compared to those of other fermions. 
This question can be interpreted to ask why the coefficient of 
the dimension five Weinberg operator~\cite{Weinberg} $(c/\Lambda)L_LL_L\Phi\Phi$ ($c$ and $\Lambda$ being dimensionless and dimensionful parameters, respectively) 
is quite small, which gives Majorana neutrino masses after replacing the Higgs field $\Phi$ by the 
vacuum expectation value (VEV). 
From the various measurements of neutrino oscillations, typically $(c/\Lambda)$ must be the order of $10^{-14}$ GeV$^{-1}$. 

The seesaw mechanism~\cite{typeI} 
gives us the simple explanation why we have such a small coefficient, namely, 
a right-handed neutrino is quite heavy, e.g., ${\cal O}(10^{14})$ GeV with a ${\cal O}(1)$ Yukawa coupling. 
However, to test this mechanism, we need a super high energy collider, and it seems to be unrealistic. 

As an alternative scenario, there are radiative neutrino mass models originated from the Zee model~\cite{Zee}. 
In these models, the smallness of the coupling $c$ can be naturally explained by a loop suppression factor $1/(16\pi^2)^N$ in a $N$-loop model. 
As a result, we do not need to introduce super heavy particles such as the above mentioned right-handed neutrino, and thus we can test these models at collider experiments. 
In addition, in radiative neutrino mass models, some of new particles which appear in a loop diagram for the neutrino mass generation 
can be a dark matter (DM) candidate as in the model by Ma~\cite{Ma}. 
Empirically, three loop models can provide appropriate size of the coefficient $c$ with TeV scale new particles and 
order of 0.1-1 couplings. 
The model by Krauss, Nasri and Trodden~\cite{KNT} is the first proposed three loop model with DM. 
In addition to those models, it has been proposed other three loop models in Refs.~\cite{AKS,GNR,Kajiyama,Culjak}. 

In this Letter, we build a new type of a three loop neutrino mass model with DM candidates. 
To have three loop diagrams, we introduce hypercharge $Y=3/2$
isospin doublet scalar and lepton fields which include doubly and singly charged components. 
As a bonus, the anomaly in the muon $g-2$ can also be simultaneously explained by loop effects of the $Y=3/2$ particles.
Characteristic signatures at collider experiments are expected from the decay of the doubly charged scalar bosons, where they can mainly decay into the same-sign dilepton with 
the both right-handed and left- and right-handed chirality. 
This signature can be used to distinguish the other models with doubly charged scalar bosons. 

\begin{center}
\begin{table*}[t]
{\small
\hfill{}
\begin{tabular}{c||c|c|c|c||c|c|c|c}\hline\hline 
&\multicolumn{4}{c||}{Lepton Fields}&\multicolumn{4}{c}{Scalar Fields}  \\\hline 
     & $L_L^i=(\nu_L^i,e_L^i)^T$ & $ e_R^i $  &  $L_{3/2}^i=(L_{3/2}^{-i},L_{3/2}^{--i})^T$ & $N_R^i$ & $\Phi$ & $\Phi_{3/2}$  & $S^+$  & $S^0$ 
  \\\hline
$(SU(2)_L,U(1)_Y)$ & $(\bm{2},-1/2)$ & $(\bm{1},-1)$ & $(\bm{2},-3/2)$ & $(\bm{1},0)$
& $(\bm{2},1/2)$  & $(\bm{2},3/2)$ & $(\bm{1},1)$  & $(\bm{1},0)$    \\\hline
$Z_2$ & $+$ & $+$ & $-$ & $-$ & $+$ & $+$  & $-$ & $-$   
\\\hline\hline
\end{tabular}}
\hfill{}
\caption{The particle contents and charge assignments. The superscript $i$ denotes the flavor index ($i=1\text{-}3$). }
\label{tab:1}
\end{table*}
\end{center}

\section{ The Model}

We discuss a three loop neutrino mass model in the electroweak $SU(2)_L\times U(1)_Y$ gauge theory with an additional unbroken discrete $Z_2$ symmetry. 
The particle contents are shown in Table~\ref{tab:1}. 
We extend the lepton sector from the SM, i.e.,  we introduce 
vector-like lepton fields $L_{3/2}^i$ with $Y=-3/2$ and right-handed neutrinos $N_R^i$ ($i=1$-3).  
The scalar sector is composed of isospin doublet fields with $Y=3/2$ $(\Phi_{3/2})$ and with $Y=1/2$ $(\Phi)$, 
and two singlets $S^+$ with $Y=1$ and $S^0$ with $Y=0$. 
For simplicity, we take $S^0$ to be a real field. 

The $Z_2$ symmetry plays three important roles in our model. 
First, it forbids the $\overline{L_L} \epsilon \Phi^* N_R$ term ($\epsilon$ being the 2 by 2 anti-symmetric matrix)
which provides the tree level Dirac neutrino mass term. 
Second, it also forbids the $\overline{L_{3/2}}\epsilon\Phi^* e_R^{}$ 
term which gives the tree level mixing between the exotic charged leptons and the SM charged-leptons. 
Finally, the $Z_2$ symmetry guarantees the stability of the lightest neutral $Z_2$ odd particle, i.e., $N_R$ or $S^0$, so that 
it can be a candidate of DM. 

In this setup, the lepton sector Lagrangian is given by 
\begin{align}
-\mathcal{L}_{\text{lep}}
&= \frac{1}{2}M_N^i \overline{N^{ic}_R} N_R^i +M_{3/2}^i \overline{(L_{3/2}^i)_L} (L_{3/2}^i)_R + {\rm h.c.} \nn \\
&+y_{\text{SM}}^{ij}\, \overline{L_L^i} \Phi e_{R}^j+y_R^{ij} \, \overline{e_R^{ic}} N_R^j S^+ + {\rm h.c.} \nn \\
&+ f_L^{ij}\,\overline{L^i_L}(L_{3/2}^j)_R S^+  
 + f_R^{ij}\, \overline{(L_{3/2}^{ic})_R}\epsilon \Phi_{3/2} N_R^j   
+ f_{LR}^{ij}\, \overline{(L_{3/2}^i)_L}\epsilon \Phi_{3/2}^* N_R^j + {\rm h.c.}, 
\end{align}
where the first and second terms in the first line correspond to the Majorana masses for $N_R^i$ and the Dirac masses for $L_{3/2}^i$, respectively. 
In the second line, $y_{\text{SM}}^{}$ represents the usual Yukawa matrix which provides the masses of the SM charged-leptons. 
All the other terms give new Yukawa interactions which are necessary to obtain left-handed Majorana neutrino masses excepted for the $y_R$ term. 

The most general Higgs potential is given by 
\begin{align}
V_{\text{Higgs}}
&=
\mu^2|\Phi|^2 +\mu_{3/2}^2|\Phi_{3/2}|^2  + \mu_\pm^2 |S^+|^2 + \frac{1}{2}\mu_0^2(S^0)^2 \nn\\
&+ \lambda_1|\Phi|^4
+ \lambda_2|\Phi_{3/2}|^4 
+ \lambda_3|S^+|^4 
+ \lambda_4(S_0)^4 \nn\\
&+\rho_1|\Phi|^2|\Phi_{3/2}|^2
 +\rho_2|\Phi|^2|S^+|^2 
 +\rho_3|\Phi|^2(S^0)^2 \nn \\
& +\rho_4|\Phi_{3/2}|^2|S^+|^2 + \rho_5|\Phi_{3/2}|^2(S^0)^2
 + \rho_6|S^+|^2(S^0)^2
\nn\\
&+  \sigma|\Phi^\dag\Phi_{3/2}|^2  
+ [\kappa\Phi_{3/2}^\dag \Phi  S^+ S^0
+\xi \Phi_{3/2}^T\epsilon \Phi (S^-)^2  + {\rm h.c.} ], 
\label{main-lag}
\end{align}
where the scalar doublet fields $\Phi$ and $\Phi_{3/2}$ can be parameterized as 
\begin{align}
\Phi &=\left[
\begin{array}{c}
G^+\\
\frac{1}{\sqrt{2}}(v+h+iG^0)
\end{array}\right],\quad
\Phi_{3/2} =\left[
\begin{array}{c}
\phi_{3/2}^{++}\\
\phi_{3/2}^{+}
\end{array}\right], 
\end{align}
with $v~(\simeq 246$ GeV) being the VEV, 
and $G^\pm$ and $G^0$ being the Nambu-Goldstone bosons which are absorbed into the longitudinal components of $W^\pm$ and $Z$ bosons, respectively. 
In our model, $h$ corresponds to the discovered Higgs boson with the mass of about 125 GeV. 
We note that the singly-charged scalar bosons $(S^\pm,\phi_{3/2}^\pm)$ do not mix with each other because of the $Z_2$ parity. 
Among the parameters in the potential, $\kappa$ and $\xi$ terms play a crucial role for the neutrino mass generation and 
the decay of $\Phi_{3/2}^{\pm\pm}$, respectively, as we shall see below. 

The squared masses $m_{\varphi}^2$ of physical scalar bosons $\varphi(=\Phi_{3/2}^{\pm\pm},~\Phi_{3/2}^{\pm\pm},~S^\pm,~S^0,~h)$ 
are then simply given without introducing mixing angles as follows
\begin{align}
&m_h^2 = 2\lambda v^2, \notag\\
&m_{\Phi_{3/2}^{\pm\pm}}^2 = \mu_{3/2}^2 + \frac{v^2}{2}\rho_1,\quad
m_{\Phi_{3/2}^{\pm}}^2 = \mu_{3/2}^2 + \frac{v^2}{2}(\rho_1+\sigma),\notag\\
&m_{S^\pm}^2 = \mu_\pm^2 + \frac{v^2}{2}\rho_2,\quad
m_{S^0}^2 = \mu_0^2 + v^2\rho_3. 
\end{align}
We here note that in our model, only the Higgs doublet field $\Phi$ gets the non-zero VEV, so that 
the masses of the SM weak gauge bosons as well as charged fermions are given exactly the same form as those in the SM.


\begin{figure}[t]
\begin{center}
\includegraphics[scale=0.8]{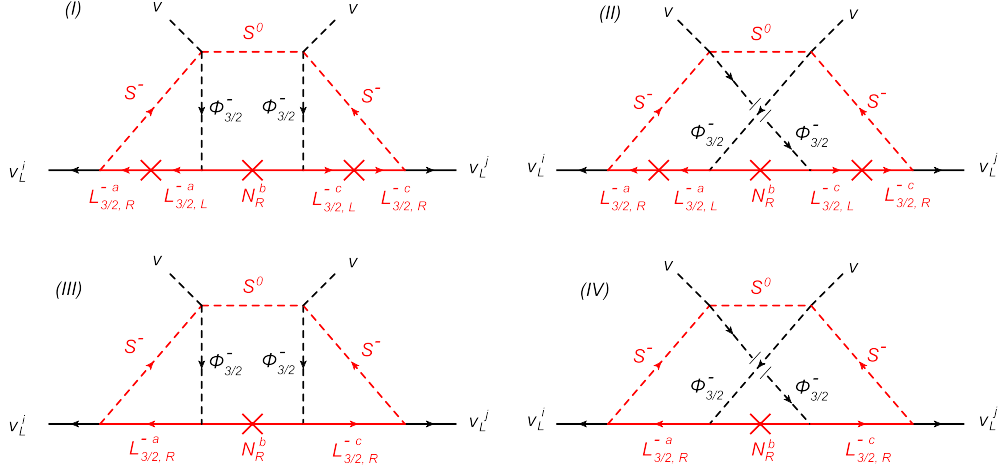}
   \caption{Three-loop neutrino mass diagrams. The red lines denote the $Z_2$ odd particles.  }
   \label{3loop}
\end{center}
\end{figure}

\section{Dark Matter}

The lightest neutral $Z_2$ odd particle can be a DM candidate; i.e., $N_R^i$ or $S^0$. 
Let us first suppose $N_R^1$ to be the lightest particle among the $Z_2$ odd particles.  
In this case, only the $t$- and $u$-channel processes $N_R^1 N_R^1 \to \ell^+\ell^-$ ($\ell^\pm$ being the SM charged lepton) 
contributes to the DM annihilation via the $S^\pm$ mediation at the tree level. 
This gives the p-wave amplitude, and the cross section is calculated by
\begin{align}
\sigma{v_{\rm rel}}=\frac{|y_R^{\ell 1}|^4}{48\pi M_{N^1}^2}
\frac{1+x^2}{\left(1+x\right)^4}v_{\rm rel}^2,
\label{eq:p-wave}
\end{align}
where $x\equiv m_{S^\pm}^2/M_{N^1}^2$ and $v_{\rm rel}$ is the relative velocity of DM. 
We note that there is no s-wave contribution due to the chiral suppression. 
To reproduce the observed relic density, 
the cross section given in Eq.~(\ref{eq:p-wave}) should be inside the following region
\begin{align}
\sigma{v_{\rm rel}}=(1.78\text{-}1.97)\times10^{-9}\ {\rm GeV^{-2}},
\end{align}
at the $2\sigma$ level. This can be satisfied by taking $y_R^{\ell 1}$ to be ${\cal O}$(1) with the mass of DM of ${\cal O}$(0.1-1) TeV. 

Next, we give a brief comment on the bosonic DM ($S^0$) case. 
This corresponds to the so-called Higgs portal DM scenario, where DM is annihilated via the $s$-channel Higgs boson $h$ mediation.
In Ref.~\cite{HiggsPotal}, it has been found that 
the DM mass of about 500 GeV can satisfies the relic density without conflicting the constraint from the 
direct searches such as the LUX experiment~\cite{LUX}.  
In addition to this solution, when the DM mass is around the half of the Higgs boson $h$ mass, i.e., $\sim 63$ GeV, 
both the relic density cross section and the direct search experiments can be satisfied using the $h$ pole effect.  

\section{Neutrino Mass}

In our model setup shown in Table~\ref{tab:1}, 
the tree level contribution to the left-handed neutrino masses, i.e., via the type-I seesaw mechanism 
is forbidden by the $Z_2$ symmetry. 
In addition, there are no one- and two-loop diagrams, where 
a systematic classification for one- and two-loop diagrams for Majorana neutrino masses 
has been done in Refs.~\cite{loop1,loop2}.  
As a result, the leading order contribution to the neutrino masses is given at the three loop level as depicted in Fig.~\ref{3loop}, where 
the topology of these diagrams is the same as neutrino mass diagrams in Ref.~\cite{AKS}.

There are four types of contributions to the neutrino masses which are denoted by the diagrams (I), (II), (III) and (IV). 
The contributions of (I) and (II) pick up two Dirac masses of $L_{3/2}$, while those of (III) and (IV) do not. 
We note that the diagram (I) gives a different contribution from 
the diagram (II), because of the difference of the loop momentum flow. 
Similarly, diagrams (III) and (IV) provide a different results with each other. 

The mass matrix for the neutrinos is then separately calculated by the four parts:
\begin{align}
&({\cal M}_\nu)_{ij} = \frac{1}{(16\pi^2)^3}\frac{\kappa^2 v^2}{2 M_{\text{max}}^2}\sum_{a,b,c=1}^3 
\times \Big[
({\cal M}_\nu^{\text{I}})_{ij}^{abc} + ({\cal M}_\nu^{\text{II}})_{ij}^{abc} + ({\cal M}_\nu^{\text{III}})_{ij}^{abc} + ({\cal M}_\nu^{\text{IV}})_{ij}^{abc}\Big], \label{mnu}
\end{align}
where $M_{\text{max}}$ denotes the largest value of the masses among $\Phi_{3/2}^\pm$, $L_{3/2}^{\pm a}$, $N_R^b$, $S^\pm$ and $S^0$. 
$({\cal M}_\nu^{\text{I-IV}})_{ij}^{abc}$ correspond to the contributions from the diagram I-IV, respectively. 
Each contribution is calculated by
\begin{align}
({\cal M}_\nu^{\text{I}})_{ij}
&=\frac{
f_L^{ia} M_{3/2}^a f_{LR}^{ab}M_N^b  f_{LR}^{cb} M_{3/2}^c f_L^{jc}
}
{M_{\text{max}}^2}
F_{\text{I}}, \label{f1}\\
({\cal M}_\nu^{\text{II}})_{ij}
&=\frac{
f_L^{ia} M_{3/2}^a f_{LR}^{ab}M_N^b f_{LR}^{cb} M_{3/2}^cf_L^{jc}
}
{M_{\text{max}}^2}
F_{\text{II}}, \label{f2}\\
({\cal M}_\nu^{\text{III}})_{ij}
&=f_L^{ia}(f_{R}^\dagger)^{ab}M_N^b (f_R^\dagger)^{bc} f_L^{cj}F_{\text{III}}, \label{f3}\\
({\cal M}_\nu^{\text{IV}})_{ij}
&=c_{ijac}f_L^{ia} (f_R^\dagger)^{ab}M_N^b (f_R^\dagger)^{bc} f_L^{cj}  F_{\text{IV}}\label{f4}, 
\end{align}
where $F_{\text{I-VI}}$ denote the loop functions for each diagram. 
Although the loop functions are determined by fixing all the mass parameters of the particles running in the loop diagram, 
their typical values, i.e., $ {\cal O}(0.1)$ are not so sensitive to these mass parameters.  

The neutrino mass matrix is parametrized by   
\begin{align}
({\cal M}_\nu)_{ij}= (U_{\rm PMNS} \, m^{\text{diag}}_\nu \, U_{\rm PMNS}^T)_{ij},
\end{align}
where $m^{\text{diag}}_\nu\equiv \text{diag}(m_1,m_2,m_3)$ is the diagonalized neutrino mass matrix with three mass eigenvalues $m_1$, $m_2$ and $m_3$.  
The Pontecorvo-Maki-Nakagawa-Sakata matrix~\cite{Maki:1962mu, Pontecorvo:1967fh} $U_{\rm PMNS}$
is the 3$\times$3 mixing matrix to diagonalize the neutrino mass matrix. 
A global fit on the recent neutrino oscillation data~\cite{Forero:2014bxa} provides their two squared mass differences 
and three mixing angles of $U_{\rm PMNS}$, depending on the mass ordering. 
The total neutrino mass $\sum_k^3 m_k$ is constrained from the experiment by Planck Collaboration~\cite{Ade:2013zuv}, which suggests $\sum_k^3 m_k<0.23$ eV at 95 \% CL.
We expect that observed three mixing angles can easily be reproduced by the Yukawa couplings $f_L$, $f_R$ and $f_{LR}$ as seen in Eqs.~(\ref{f1})-(\ref{f4}).
In particular, the case with diagonal $f_{L}$ and $f_R$ couplings is favored to avoid lepton flavor violating processes, and 
even in such a case, the mixing can be accommodated by $f_{LR}$.

We here estimate typical values of the parameters 
that appear in the above neutrino mass formulae. 
Assuming that each of ${\cal M}_\nu^{\text{I-IV}}$ is the order of 0.1 eV 
as required by the magnitude of the observed neutrino masses, we obtain the following relation for the contribution, e.g., from ${\cal M}_\nu^{\text{I}}$:
\begin{align}
\frac{1}{(16\pi^2)^3}\times (\kappa f_Lf_{LR})^2F_I^{}\times \frac{v^2M_{3/2}^2M_N}{M_{\text{max}}^4} = {\cal O}(0.1)~\text{eV}, \label{est}
\end{align}
where the flavor indices are omitted here. 
When we take $M_{\text{max}}\simeq M_{3/2}\simeq M_{N}$, and $F_I^{}\simeq 0.1$, Eq.~(\ref{est}) can be rewritten as 
\begin{align}
(\kappa f_L f_{LR})^2 \frac{v}{M_{\text{max}}}  ={\cal O}(10^{-5}). 
\end{align}
In the case of $M_{\text{max}}={\cal O}(1)$ TeV, the factor $(\kappa f_L f_{LR})^2$ should be the order of ${\cal O}(10^{-4})$ which can be realized by taking 
each of $\kappa$, $f_L$ and $f_{LR}$ couplings to be in the range of 0.1 to 1. 
Therefore, tiny neutrino masses can be $naturally$ explained by introducing neither super heavy particles nor
quite small coupling constants as compared to the gauge coupling constants.

\begin{figure}[t]
\begin{center}
\includegraphics[scale=0.7]{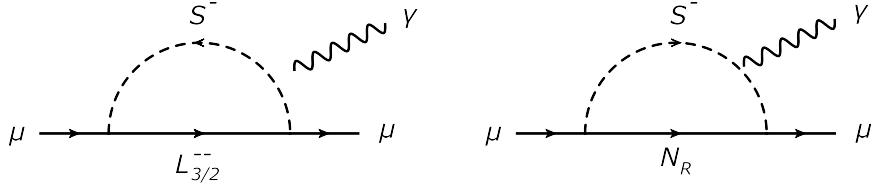}
   \caption{New contributions to the muon $g-2$.  In the left diagram, the photon can be emitted from either $L_{3/2}^{--}$ or $S^-$.  }
   \label{gminus2}
\end{center}
\end{figure}

\section{Muon anomalous magnetic moment}

In our model, 
it is possible to explain the anomaly in the muon anomalous magnetic moment (muon $g-2$) as a bonus. 
It has been known that there is a discrepancy between the SM prediction and the measured value at 
Brookhaven National Laboratory. 
The observed value is given by~\cite{bennett}
\begin{align}
a^{\rm exp}_{\mu}=11 659 208.0(6.3)\times 10^{-10}, \notag
\end{align}
and the difference between the above value and the SM prediction ($\Delta a_{\mu}\equiv a^{\rm exp}_{\mu}-a^{\rm SM}_{\mu}$) was calculated as 
\begin{align}
&\Delta a_{\mu}=(29.0 \pm 9.0)\times 10^{-10}~\text{\cite{discrepancy1}},~~\text{and} \label{dev1} \\
&\Delta a_{\mu}=(33.5 \pm 8.2)\times 10^{-10}~\text{\cite{discrepancy2}}. \label{dev2}
\end{align}
The above results given in Eqs. (\ref{dev1}) and (\ref{dev2}) correspond
to $3.2\sigma$ and $4.1\sigma$ deviations, respectively. 

There are new contributions to $\Delta a_\mu$ as shown in Fig.~\ref{gminus2}. 
These are calculated as
\begin{align}
\Delta a_\mu &= \frac{m_\mu^2}{16\pi^2}
\Bigg\{|f_L^{\mu\mu}|^2\left[
G\left(M_{3/2}^\mu,m_{S^\pm}\right) + 2G\left(m_{S^\pm},M_{3/2}^\mu\right)\right] 
 -|y_R^{\mu\mu}|^2 G\left(M_N^\mu,m_{S^\pm}\right)\Bigg\}, 
\label{eq:muon-g-2}
\end{align}
where
\begin{align}
G(m_1,m_2) =
& \frac{2 m_1^6+3m_1^4 m_2^2 - 6m_1^2 m_2^4 + m_2^6 -6m_1^4 m_2^2\ln\frac{m_1^2}{m_2^2}}{6(m_1^2-m_2^2)^4}. 
\end{align}
In this loop function, we neglect the muon mass dependence. 
In Eq.~(\ref{eq:muon-g-2}), we take both $f_L$ and $y_R$ couplings to be a diagonal form in order to avoid constraints from lepton flavor violating processes such as 
$\mu \to e\gamma$~\footnote{There are $\ell \to 3\ell$ type lepton flavor violating processes via box diagrams~\cite{Toma,Nishiwaki}. However, these contributions 
can also be avoided by taking diagonal couplings of $y_R$ and $f_L$. }. 
Thus, only $L_{3/2}^{--\,\mu}$ and $N_R^{\mu}$ contribute to the muon $g-2$. 

\begin{figure}[t]
\begin{center}
\includegraphics[scale=0.25]{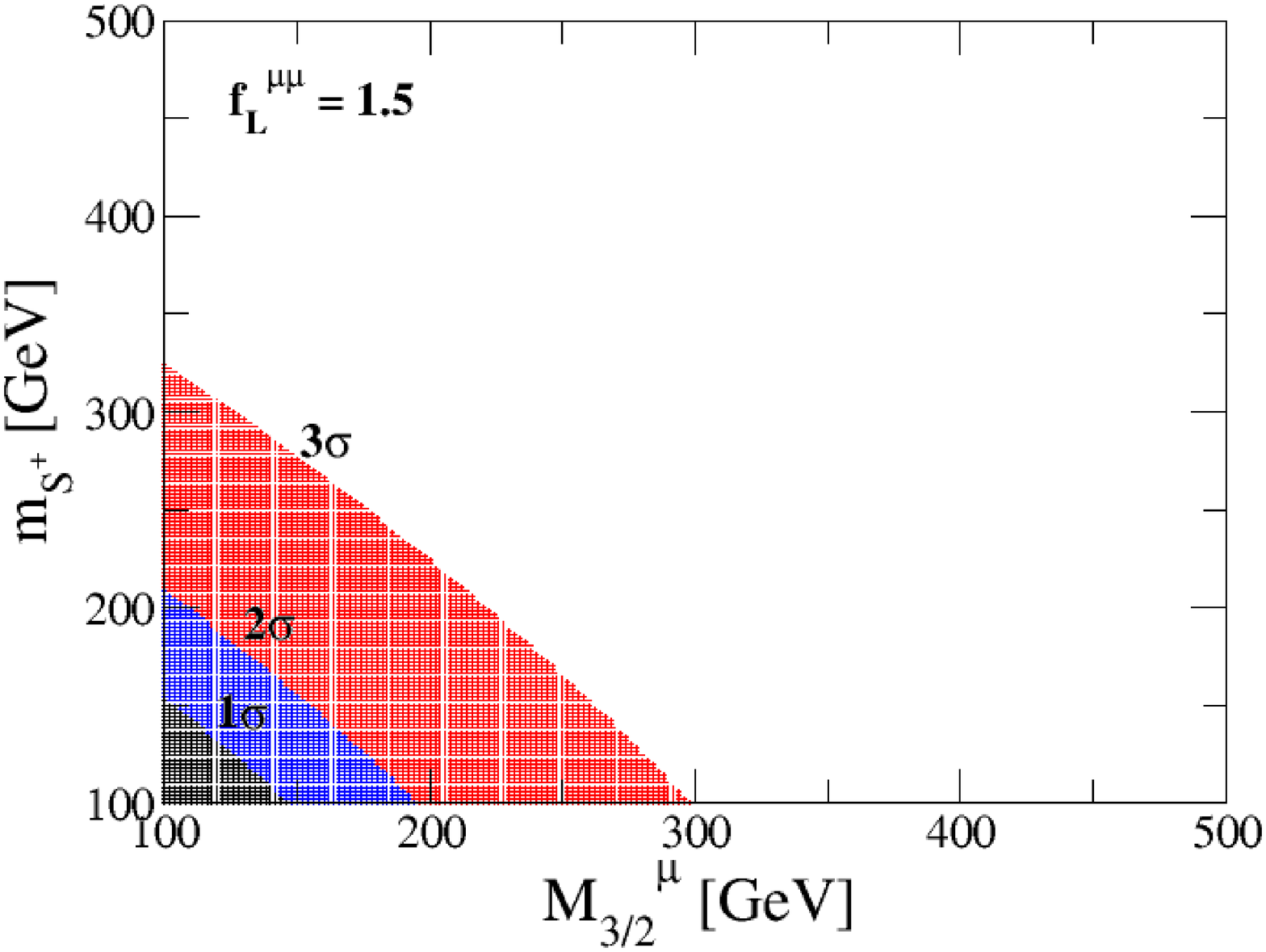}
\hspace{3mm}
\includegraphics[scale=0.25]{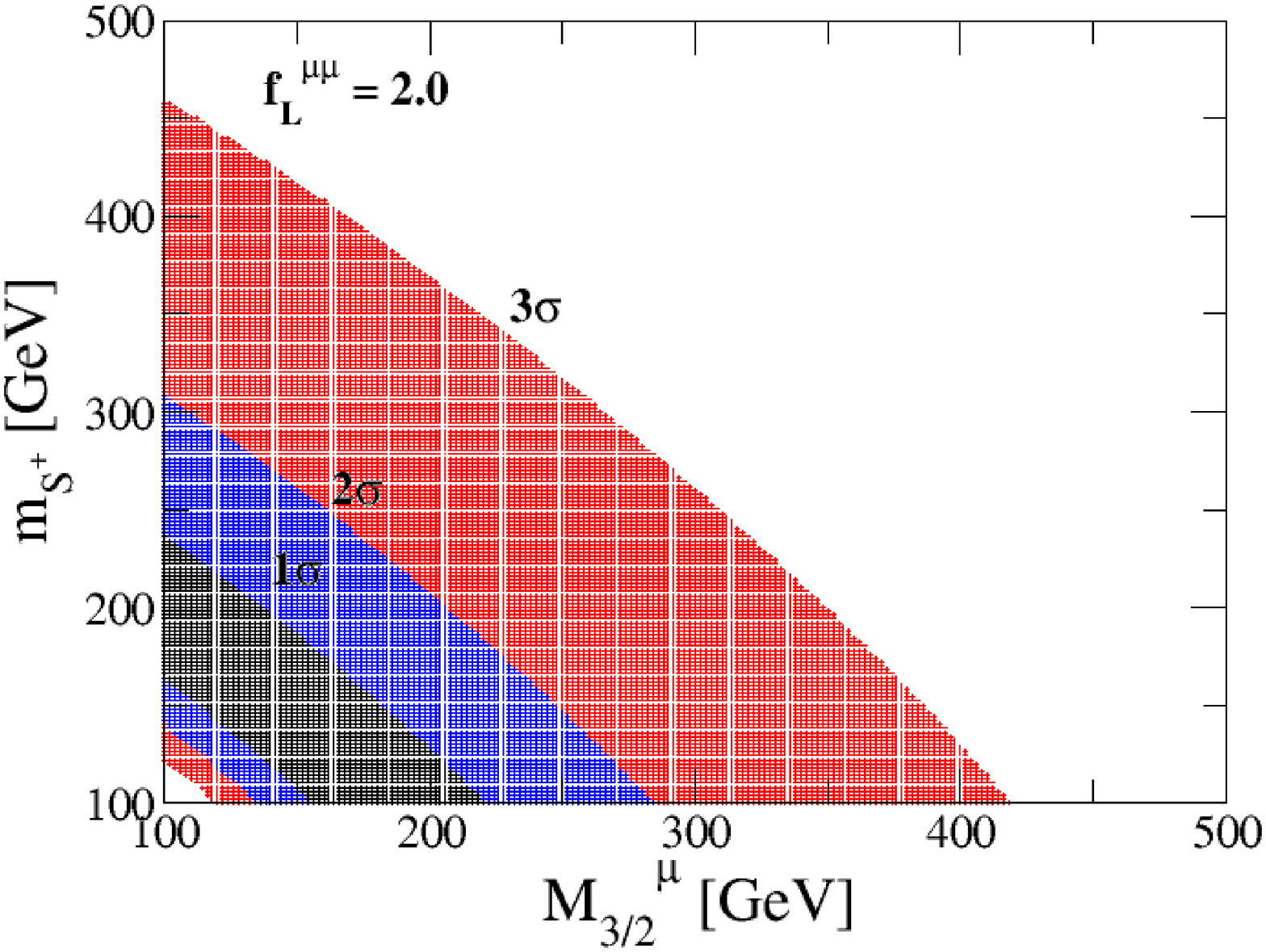} \\
\includegraphics[scale=0.25]{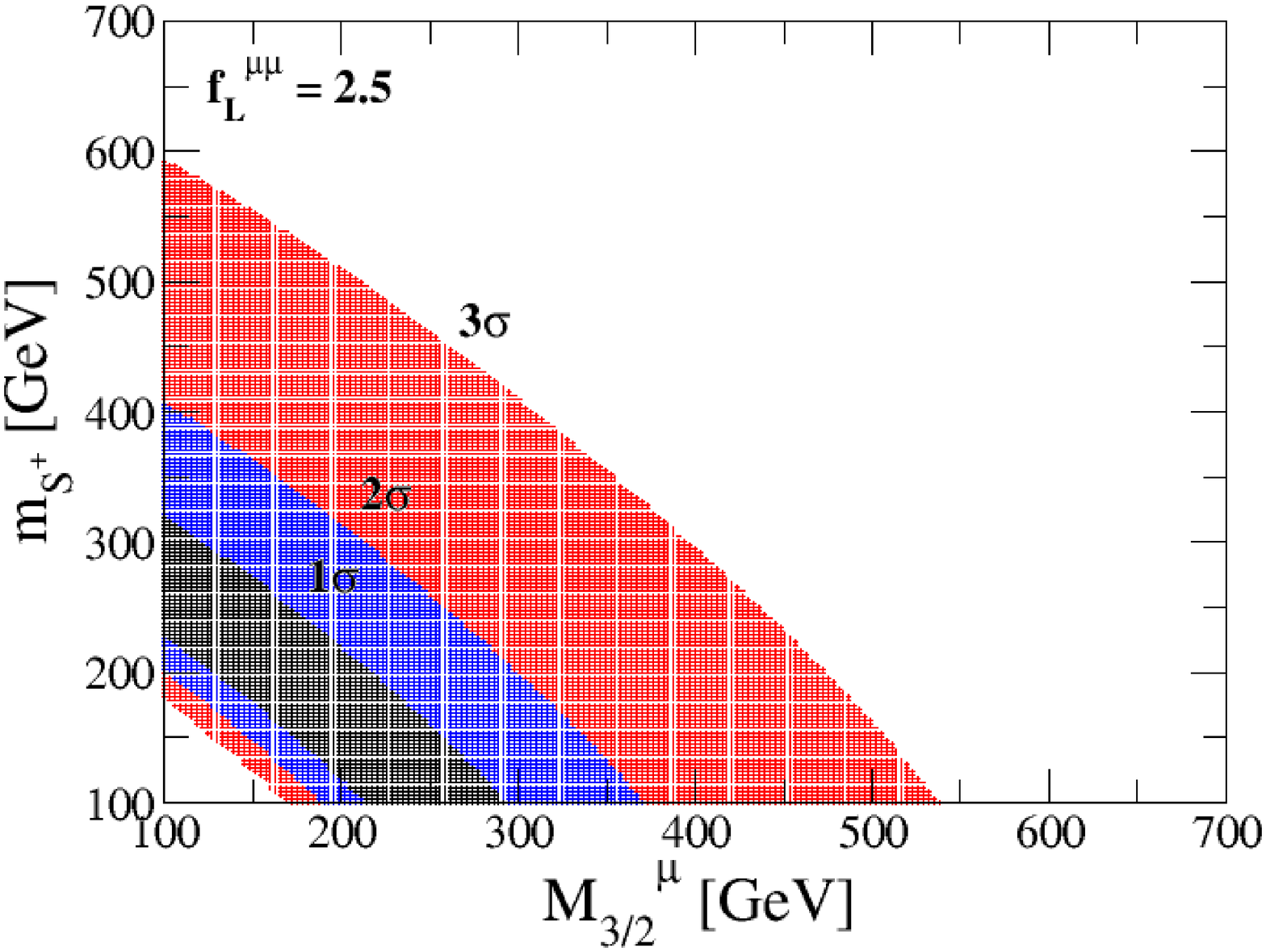}
\hspace{3mm}
\includegraphics[scale=0.25]{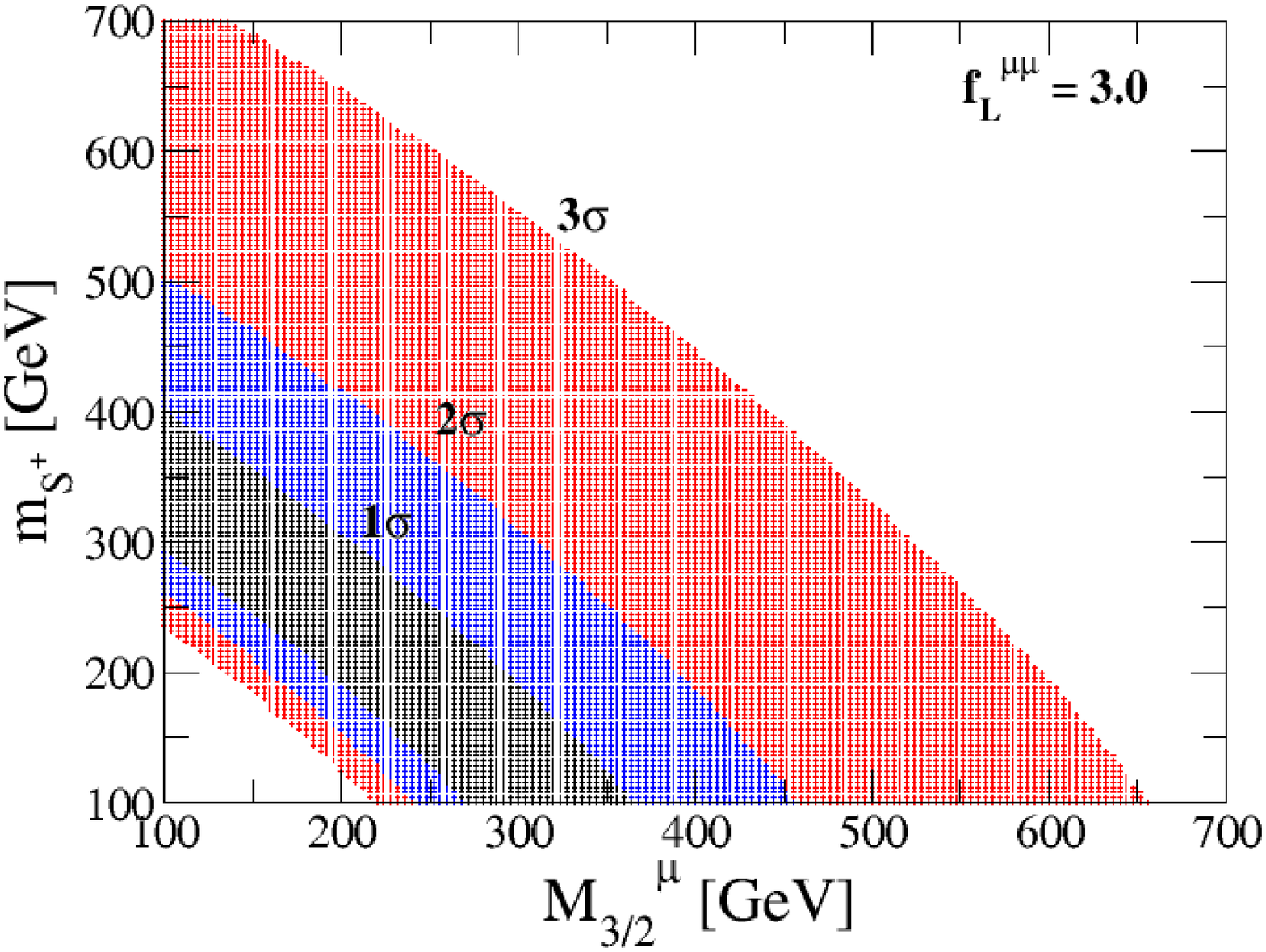} 
   \caption{Parameter region which can explain the discrepancy of the muon $g-2$ on the $M_{3/2}^\mu$-$m_{S^\pm}^{}$ plane. 
The black, blue and red shaded regions give the prediction of $\Delta a_\mu$ within the 1$\sigma$, 2$\sigma$ and $3\sigma$ error given in Eq.~(\ref{dev2}).  
The value of $f_L^{\mu\mu}$ is fixed to be 1.5, 2.0, 2.5 and 3.0 in the upper-left, upper-right, lower-left and lower-right panels, respectively. 
In all the plots, we take $y_R^{\mu\mu}=0$. }
   \label{figg-2}
\end{center}
\end{figure}

In Fig.~\ref{figg-2}, we show the parameter region which can explain the anomaly of the muon $g-2$ on the $M_{3/2}^\mu$-$m_{S^\pm}^{}$ plane. 
We here use the discrepancy of $\Delta a_\mu$ given in Eq.~(\ref{dev2}). 
In these plots, we take $y_{R}^{\mu\mu}=0$ to remove the contribution from $N_R^\mu$ which 
always gives the distractive effect to the $L_{3/2}^{--\,\mu}$ loop contribution as seen in Eq.~(\ref{eq:muon-g-2}). 
In this case, $\Delta a_\mu$ is given as a function of $m_{S^\pm}$, $M_{3/2}^\mu$ and $f_L^{\mu\mu}$. 
The value of $f_L^{\mu\mu}$ is fixed to be 1.5, 2.0, 2.5 and 3.0 in the upper-left, upper-right, lower-left and lower-right panels, respectively.
The parameter region indicated by black, blue and red shaded regions gives the prediction of $\Delta a_\mu$ within the 1$\sigma$, 2$\sigma$ and $3\sigma$ error. 
We find that the $2\sigma$ allowed region is obtained by taking $m_{S^\pm}+M_{3/2}^\mu \lesssim 400$, 600, 750 and 950 GeV for $f_L^{\mu\mu}=1.5$, 2.0, 2.5 and 3.0, respectively.

\section{Collider phenomenology}

\begin{figure}[!t]
\begin{center}
\includegraphics[scale=0.7]{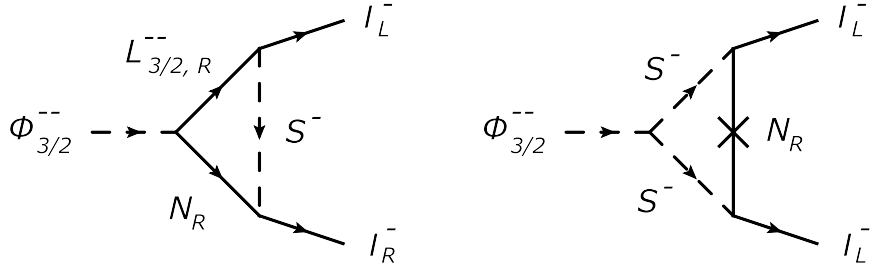}
   \caption{One loop induced decays of $\Phi_{3/2}^{--}$.   }
   \label{collider}
\end{center}
\end{figure}

In this section, we discuss collider signature of the new particles in our model. 
The most important signature to identify our model 
could come from the doubly charged scalar bosons $\Phi_{3/2}^{\pm\pm}$ and leptons $L_{3/2}^{\pm\pm}$.
At the LHC, these particles are produced in pair via the Drell-Yan process, i.e., 
$pp \to \Phi_{3/2}^{++}\Phi_{3/2}^{--}$ and 
$pp \to L_{3/2}^{++}L_{3/2}^{--}$, and also via associated processes with singly-charged particles, i.e., 
$pp \to \Phi_{3/2}^{\pm\pm}\Phi_{3/2}^{\mp}$ and 
$pp \to L_{3/2}^{\pm\pm}L_{3/2}^{\mp}$. 

The decay pattern of these doubly charged particles strongly depends on the mass spectrum of the new particles 
which can be roughly determined from the discussion in the previous sections and experimental constraints. 
From the discussion in Sec.~V, we found that we need to take masses of $L_{3/2}^{--\,\mu}$ and $S^\pm$ to be a few hundred GeV  
to explain the anomaly in the muon $g-2$. 
In addition, if we consider the case of the scalar DM, i.e., $S^0$, 
$m_{S^0}$ should be taken to be about 63 GeV as we explained in Sec.~III. 
Although in this case the masses of $N_R^{i}$ are not constrained so much, it should be at most ${\cal O}(1)$ TeV to reproduce the 
order of neutrino masses as we discussed in Sec.~IV.  
For the mass of $\Phi_{3/2}^{\pm\pm}$, the most important bound comes from the same-sign dilepton searches at the LHC. 
The current bound using the LHC Run-I data gives the lower limit on the mass of $\Phi_{3/2}^{\pm\pm}$ to be about 550 GeV when 
$\Phi_{3/2}^{\pm\pm}$ mainly decay into the same-sign dielectron~\cite{ATLAS}. 
Weaker bounds are obtained if the flavor of the same-sign dilepton from the decay of $\Phi_{3/2}^{\pm\pm}$ is different from the electron. 
We will discuss later how the same-sign dilepton decay of $\Phi_{3/2}^{\pm\pm}$ is realized in our model. 
Regarding to the doubly charged leptons $L_{3/2}^{\pm\pm}$, 
we cannot apply the similar bound for the mass of $\Phi_{3/2}^{\pm\pm}$ to that of $L_{3/2}^{\pm\pm}$, 
because their decay products must include the DM in the final state due to
those $Z_2$ odd property. 
Thus, the typical final state of the decay of $L_{3/2}^{\pm\pm}$ is the same-sign dilepton plus missing energy, and 
the mass bound must be weaker than that of $\Phi_{3/2}^{\pm\pm}$. 
The study for the collider phenomenology of doubly charged leptons have been studied in Ref.~\cite{dcl}. 

Under this configuration, let us consider the following mass spectrum: 
\begin{align}
M_N~(\sim {\cal O}(1)\,\text{TeV}) > m_{\Phi_{3/2}}^{} > M_{3/2}\gtrsim m_{S^\pm} > m_{S^0}~(\sim 63\,\text{GeV}), 
\end{align}
where $m_{\Phi_{3/2}}^{} \equiv m_{\Phi_{3/2}^{\pm\pm}}(=m_{\Phi_{3/2}^{\pm}})$\footnote{If there is a non-zero mass difference between 
$m_{\Phi_{3/2}^{\pm\pm}}^{}$ and $m_{\Phi_{3/2}^{\pm}}^{}$, the cascade decay, i.e., $\Phi_{3/2}^{\pm\pm}\to \Phi_{3/2}^{\pm} W^{\pm *}$ is possible.  
In Ref.~\cite{Aoki}, the collider phenomenology for such a cascade decaying doubly charged scalar bosons has been discussed. }, and 
we take the flavor universal masses $M_{3/2}\equiv M_{3/2}^{i=1,2,3}$ and $M_N\equiv M_N^{i=1,2,3}$ for simplicity. 
In this mass spectrum, the following decay modes can be considered:
\begin{align}
&\Phi_{3/2}^{\pm\pm} \to S^\pm S^\pm \quad (\text{via}~\xi), \quad \Phi_{3/2}^{\pm\pm} \to \ell^\pm \ell^\pm \quad (\text{loop induced}), \label{decay1}\\
& L_{3/2}^{\pm\pm} \to S^\pm \ell_L^\pm  \quad   (\text{via}~f_L), 
\end{align}
with the subsequent decay\footnote{The $\Phi_{3/2}^\pm \to \ell \nu$ decay is given via similar diagrams as shown in Fig.~\ref{collider}. } 
of $S^\pm \to \Phi_{3/2}^{\pm *} S^0 \to \ell^\pm \nu S^0$ via the $\kappa$ term in the potential. 
In (\ref{decay1}), the loop induced decay into the same-sign dilepton is driven via the diagram shown in Fig.~\ref{collider}. 
This process can also be understood by the following dimension five operators with a typical mass scale of our model $M$: 
\begin{align}
\frac{1}{M} \overline{L_L}\Phi_{3/2} \partial_\mu \gamma^\mu e_R^{},\quad 
\frac{1}{M} \overline{e_R^c}e_R^{}\Phi_{3/2} \epsilon \Phi, 
\end{align}
where the first (second) operator induces the left (right) decay process of Fig.~\ref{collider}. 
The important point here is that $\Phi_{3/2}^{\pm\pm}$ can decay into 
the same-sign dilepton with its chirality of both right-handed or left- and  right-handed. 
This feature cannot be seen in other doubly charged scalar bosons from the other models as follows. 
First, 
doubly charged scalar bosons from an isospin singlet with $Y=2$, which are introduced in the two loop neutrino mass model the so-called Zee-Babu model~\cite{Zee-Babu}, 
can decay into the same-sign dilepton with only right-handed. 
Next, doubly charged scalar bosons from an isospin triplet with $Y=1$, which is introduced in the Higgs triplet model (HTM)~\cite{typeII}, 
can decay into the same-sign dilepton with only left-handed.  
Therefore, by measuring the chirality of the same-sign dilepton event, our model can be distinguished from the other models with 
doubly charged scalar bosons. 
In Ref.~\cite{Sugiyama}, it has been clarified that the same-sign dilepton signals 
with a different chirality can be distinguished by using distributions of decay products of tau leptons. 

Finally, we comment on the signal and background events in our model. 
If the same-sign dilepton decay of $\Phi_{3/2}^{\pm\pm}$ is dominant which can be realized by a smaller value of the $\xi$ coupling in the potential, 
the expected collider signature is 
$pp \to Z^*/\gamma^* \to \Phi_{3/2}^{++}\Phi_{3/2}^{--}\to \ell^+\ell^+\ell^-\ell^-$. 
Dedicated studies of the signal and background analysis for the four lepton events has been done in Refs.~\cite{4lep1,4lep2,4lep3}. 
The SM background processes $ZZ$, $WZ$, $t\bar{t}$, $Zbb$, $Ztt$ and $Wtt$ have been taken into account in Ref.~\cite{4lep3} in the detector level analysis. 
Among them, the largest background cross section comes from the $ZZ$ and $Ztt$ processes, 
but they can be significantly 
reduced by taking the invariant mass cut of the $\ell^+\ell^-$ system $M_{\ell^+\ell^-}$, i.e., 
the event with $|M_{\ell^+\ell^-}-m_Z^{}|\leq 10$ GeV is excluded~\cite{4lep3}. 
As a result of the background separation from the signal, the $5\sigma$ discovery reach for doubly charged Higgs bosons 
in the HTM has been presented in~\cite{4lep3}, where the required integrated luminosity is 6 fb$^{-1}$ (60 fb$^{-1}$)
when the mass of doubly charged Higgs bosons are taken to be 500 (700) GeV at the 14 TeV LHC. 
In our model, the production cross section of $\Phi_{3/2}^{\pm\pm}$ is smaller than that in the HTM because of the 
difference of the doubly charged Higgs boson coupling to the $Z$ boson. 
For example, if we take the mass of doubly charged Higgs bosons to be 500 (700) GeV, we obtain the pair production cross section 
of 1 (0.15) fb and 2 (0.3) fb in our model and in the HTM~\cite{su}, respectively. 
However, the same signal and background analysis given in~\cite{4lep3} can be applied in our case, so that 
we expect to obtain the 5$\sigma$ discovery reach for $\Phi_{3/2}^{\pm\pm}$ by requiring the roughly 2 times larger integrated luminosity than the case in the HTM with the same mass of 
the doubly charged Higgs bosons (due to about the 50\% reduction of the cross section as compared to the HTM) at the 14 TeV LHC.

\section{Conclusions}

We have constructed a three-loop induced neutrino mass model with the additional unbroken discrete $Z_2$ symmetry. 
The lightest neutral $Z_2$ odd particle, i.e., the right-handed neutrino $N_R$ or the real singlet scalar boson $S^0$ can be a DM candidate. 
We have shown that the neutrino masses can be naturally explained by the order 1 TeV new particles with the order 0.1-1 couplings due to the three-loop suppression. 
At the same time, the anomaly of the muon $g-2$ can also be explained by the loop effect of the doubly charged leptons $L_{3/2}^{\pm\pm}$. 
Regarding to the collider phenomenology, we have discussed the signature from doubly charged scalar bosons $\Phi_{3/2}^{\pm\pm}$ and leptons $L_{3/2}^{\pm\pm}$. 
In particular, we have emphasized that the loop induced decay of $\Phi_{3/2}^{\pm\pm}$ into the same-sign dilepton $\ell_R^\pm \ell_R^\pm$ and $\ell^\pm_L \ell^\pm_R$ can be the important 
signature to distinguish our model from the other models with doubly charged scalar bosons. 
\\\\
{\it Acknowledgments}

H.O. expresses his sincere gratitude toward all the KIAS members, Korean cordial persons, foods, culture, weather, and all the other things.
K.Y. is supported by JSPS postdoctoral fellowships for research abroad.

\end{document}